# Resolving competing distortions in $Ca_{0.4}Sr_{0.6}TiO_3$ using complementary electron and X-ray techniques.

Robin Sjökvist[1*], Catriona A. Crawford[1*], Richard Beanland[1†] and Mark S. Senn[2‡]

[1]Department of Physics, University of Warwick, Coventry CV4 7AL, United Kingdom
[2]Department of Chemistry, University of Warwick, Coventry, CV4 7AL, United Kingdom

**ABSTRACT**. We present a study of the perovskite $Ca_{0.4}Sr_{0.6}TiO_3$ using variable temperature transmission electron microscopy (TEM) and powder X-ray diffraction (PXRD). At room temperature and below, PXRD shows that the material adopts an orthorhombic $Pbcm$ structure analogous to the P-phase of $NaNbO_3$. Above 380 K the material transforms to a tetragonal $I4/mcm$ phase. The structural distortions of these phases can be described as a combination of modes and order parameters associated with the $M$, $T$, $\Delta$ and $R$-points of the Brillouin zone, each of which can be associated with a different set of superstructure reflections visible in X-ray and electron diffraction patterns. For the $I4/mcm$ phase only the expected $R$-point reflections are observed in PXRD while both $M$ and $R$- reflections are observed in electron diffraction. Using $\Delta$ and $R$ dark field TEM images we show that the phase transition proceeds by a loss of coherence of $TiO_6$ octahedral tilting along the *c*-axis, leading to a microstructure of thin nanoscale platelets with a different local symmetry to the macroscopic structure. These persist well above the phase transition temperature and are probably responsible for the long-standing discrepancy between Raman spectroscopy and diffraction measurements in this materials system, as well as other secondary effects.

## I. INTRODUCTION.

Perovskite oxides in the solid solution $Ca_xSr_{1-x}TiO_3$, have been studied for several decades,[1-3] motivated by its relevance to lower mantle earth science,[4] nuclear waste,[5] and more recently as a quasi-linear dielectric.[6] It exhibits composition-dependent displacive phase transitions and a variety of properties ranging from incipient quantum ferroelectricity at compositions close to $SrTiO_3$,[7,8] paraelectric behaviour for compositions $x > 0.41$,[9] and an antiferrodistortive (AFD) structure for compositions $x \leq 0.4$.[10,11] Understanding this complex behaviour evolved as improved experimental capabilities allowed more detailed investigation,[9] summarised in the definitive work of Carpenter[12] and supported by X-ray diffraction, (XRD),[13] neutron diffraction, (ND),[14,15] and electron diffraction (ED).[13,14]

The phase diagram for $x < 0.5$ is summarised in Figure 1(a). At room temperature (RT), $Ca_xSr_{1-x}TiO_3$ exhibits four phases: cubic $Pm\bar{3}m$ $x < 0.08$; tetragonal $I4/mcm$ $0.08 < x < 0.35$; orthorhombic $Pbcm$ $0.35 < x < 0.41$; and orthorhombic $Pnma$ for $x > 0.41$. At any given composition on cooling from high temperature there are two phase transitions, first from the cubic aristotype $Pm\bar{3}m$ to tetragonal $I4/mcm$ at temperature $T_{C1}$, followed by an orthorhombic phase at $T_{C2}$, with a first-order transition that is apparent from phase coexistence and temperature hysteresis, marked by the band in Figure 1(a). The large difference for the two end members, $CaTiO_3$ $T_{C1}$ = 1650 K, and $SrTiO_3$ $T_{C1}$ = 105 K,[7] produces a very rapid, and nearly linear, suppression of both $T_{C1}$ and $T_{C2}$ with decreasing *x*.

Our study is motivated by unresolved contradictions in previous work. While the phase diagram shown in Figure 1(a) is now generally accepted,[9] prior to its establishment there were conflicting determinations of the $I4/mcm$ phase, [10,11,16–18] resulting from attempts to reconcile Raman measurements that unequivocally showed lower than tetragonal symmetry.[18,19] While there are some observations that point to an origin in some local, nanoscale structure of lower symmetry,[14,20] until now these have not yet been observed. Here, we focus on $Ca_{0.4}Sr_{0.6}TiO_3$, using both synchrotron powder X-ray Diffraction (PXRD) and Transmission Electron Microscopy (TEM) and ED over a range of temperatures to explore the $Pbcm$ to $I4/mcm$ phases and their associated transition. Using dark field TEM imaging, we observe microstructure specific to individual distortion modes and observe their evolution as the phase transition is crossed. We find the $I4/mcm$ phase contains nanoscale platelets with in-phase $TiO_6$ octahedral tilts that persist at temperatures well above the transition temperature, providing a plausible explanation for the discrepancy between Raman and XRD data in this intriguing material.

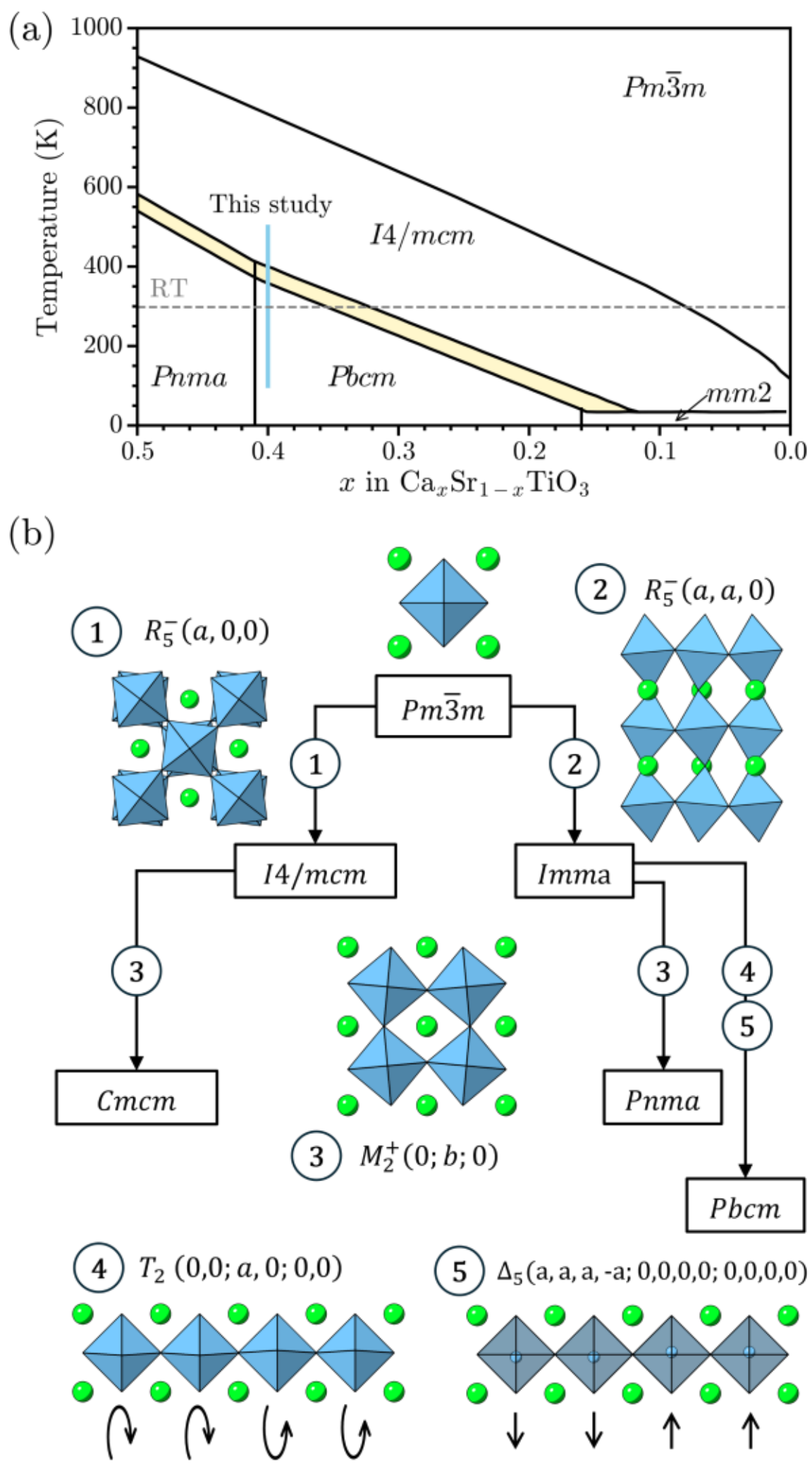

*FIG 1. (a) Phase diagram for $Ca_xSr_{1-x}TiO_3$, x < 0.5, after Carpenter [12]. Phase coexistence occurs in the region between the parallel lines at the lower bound of the tetragonal $I4/mcm$ phase, shaded yellow for clarity. (b) Schematics showing distortion modes and space group relations for most structures considered in this study.*

### A. The Structure of $Ca_xSr_{1-x}TiO_3$

Displacive phase transitions of $ABO_3$ perovskites, including those of $Ca_xSr_{1-x}TiO_3$, are produced by coordinated translations of atoms in the material, providing the lowest free energy state at a given temperature. While the most frequently mentioned distortions in perovskites are rigid-unit rotations ('tilts') of the $BO_6$ octahedra, [21,22] the range of permissible distortions is most robustly classified using irreducible representations (irreps). Each irrep can be associated with a critical point of the aristotype perovskite Brillouin zone, which links directly to the position of superstructure reflections in a diffraction pattern. The collection of symmetry adapted distortion modes associated with an irrep are often referred to as a primary or secondary order parameter (OP) in a Landau-Ginzberg description of phase transitions; the most significant OPs and their associated directions in the perovskite structure have been given by Senn & Bristowe. [23] The use of this more formal description of order parameters and associated phase transition is particularly relevant for the $Ca_xSr_{1-x}TiO_3$ $Pbcm$ phase, which cannot be described by simple in-phase or anti-phase octahedral tilting.

The OPs relevant to $Ca_xSr_{1-x}TiO_3$ are well documented[12,14] and are shown in Figure 1b. (Irrep notation here and in this work uses an A-site origin, with respect to a $Pm\bar{3}m$ parent structure. A B-site origin is also frequently used, see e.g. Senn & Bristowe[23].) The $I4/mcm$ phase only has out-of-phase $TiO_6$ octahedral tilting along one principal unit cell axis (in Glazer notation, $a^0a^0c^-$). This is captured by a single OP at the Brillouin zone $R$-point, **k** = (½ ½ ½), giving reflections appear at half odd-odd-odd (½ *ooo*) positions in diffraction patterns indexed using the (pseudo)cubic aristotype cell. This is not the only OP that can produce $R$-point reflections and is more specifically described as $R_5^-$ with order parameter direction (OPD) $(a,0,0)$.

The orthorhombic $Pnma$ phase (x > 0.41) has an $a^-a^-c^+$ structure, with two primary OPs, $R_5^-(a,a,0)$ – describing equal antiphase tilts along two principal unit cell axes, and in-phase tilting along the third axis described by the $M$-point OP $M_2^+(0;b;0)$, **k** = (½ ½ 0). Diffraction patterns thus exhibit ½ *ooo* $R$-point and ½ *eeo* $M$-point superstructure reflections.

The orthorhombic $Pbcm$ phase (0.35 < x < 0.41) also has $R_5^-(a,a,0)$ as a primary OP but octahedral tilting about the *c*-axis is changed from simple in-phase tilting to a pairwise out-of-phase (↓↓↑↑) $TiO_6$ tilting pattern over four aristotype unit cells. This is described by the OP $T_2(0,0;a,0;0,0)$ (**k** = (½ ½ ¼)), which could be expressed as $a^-a^-c^{++--}$ in Glazer notation. Additional, presumed secondary, OPs are also present, such as $\Delta_5(a,a,a,-a;0,0,0,0;0,0,0,0)$ (**k** = (0 0 ¼)), which gives antiferrodisplacive (AFD) displacements of all atomic species present in the material with the same (↓↓↑↑) pattern as the $T_2$ mode.[12] The symmetry also allows a second order Jahn-Teller (SOJT) distortion of $TiO_6$ octahedra, $M_5^-$. These different OPs produce $M$-point, $T$-point, $\Delta$-point and $R$-point superstructure reflections, which can readily be observed in PXRD and electron diffraction patterns. This structure is also observed in $NaNbO_3$ where it is known as the P-phase.[14,24] Interestingly, $NaNbO_3$ can be irreversibly transformed from $Pnma$ $a^-a^-c^{++--}$ to the $a^-a^-c^+$ $Pnma$ (T-phase) by the application of an electric field. [25] That is, while the $R_5^-$ OP is unchanged, $c$-axis octahedral tilting $T_2$ is replaced by $M_2^+$. In the case of $Ca_xSr_{1-x}TiO_3$, a similar $R_5^-$ to $T_2$ transition is found as a function of composition, at $x$ = 0.41. For our material, with $x$ = 0.40, $T_2$ wins the competition between these OPs.

*R.S. and C.A.C. contributed equally to this work
†Contact author: R.Beanland@warwick.ac.uk
‡Contact author: M.Senn@warwick.ac.uk

In our PXRD analysis below, we perform Rietveld refinements of our structural model in the basis corresponding directly to the above-described symmetry-adapted modes. When combined with high resolution synchrotron powder diffraction data, this method provides extremely precise (sub-picometre) atomic coordinates and lattice parameters for an average structure, allowing extraction of OPs that link directly to Landau-Ginzberg models of phase transformations and functional properties. Nevertheless, models that primarily involve displacement of oxygen atoms can be hard to distinguish due to the poor sensitivity to light elements, relying on the fitting of very weak peaks. This problem is compounded by Scherrer broadening, which, for radiation with wavelengths around 0.1 nm, typical for X-ray and neutron diffraction, smears out diffracted peaks from nanoscale structures, making these already weak peaks undetectable. Electron diffraction, with greater sensitivity to oxygen atoms, small wavelength (0.002 nm) and strong interaction with matter thus complements PXRD and has proved invaluable in determining the $Pnma$ phase of $Ca_xSr_{1-x}TiO_3$. Here, we extend this approach by use dark field TEM imaging of specific distortion modes to understand the link between microstructure and phase behaviour.

## II. METHODS

Pellets of $Ca_{0.4}Sr_{0.6}TiO_3$ were synthesised by pressing a ground mixture of stoichiometric $CaCO_3$, $SrCO_3$ and $TiO_2$ powders in a 5 tonne die press (Specac). The pressed pellets were calcined in alumina crucibles (heating rate of 10 °C per minute), initially at 900 °C for 10 hours and subsequently 3 cycles of 1300 °C for 12 hours each, with intermediate re-grinding and pelletising between each firing. No impurity phases were evident in the synchrotron powder diffraction data and unit cell volume is consistent with previous reports in the literature.[12]

For high resolution PXRD measurements, the sample was ground and filled into a borosilicate capillary of 0.3 mm internal diameter. Data were collected at beamline I11, Diamond Light Source (DLS) using 15 keV radiation (λ = 0.82476(1) Å, determined against a Si standard) collected with a MYTHEN detector. VT data were collected by sweeping between 100 K to 500 K with a 6 K min$^{-1}$ ramp rate. Each data collection lasted about 45 seconds and therefore encompass a temperature range of about 4.5 K. Temperatures were controlled using an Oxford Cryosystems Cryostream. Rietveld refinements on powder diffraction data were performed in TOPAS Academic Version 6 using an integrated jEdit text editor.[26,27] The background was modelled by an 8-term Chebyshev function, and peaks were modelled using a Thompson-Cox peak shape. Structural refinements of atomic positions were performed using a symmetry-motivated basis with respect to the cubic aristotype, as parameterised by ISODISTORT.[28,29] Lattice parameters were refined using symmetry-adapted strains and were parameterised in a similar manner.

For the $Pbcm$ phase, only distortion modes transforming as the irreps $T_2$ (1 mode), $\Delta_5$ (5), $R_5^-$ (1), and $M_5^-$ (3) along with symmetry adapted strains $\Gamma_1^+$ (1) and $\Gamma_3^+$ (a,-1.732a) (1) were refined. All other irreps and strains were small and fixed at zero. Above 380 K, two $I4/mcm$ models were required. All symmetry adapted distortions ($R_5^-$) and strains ($\Gamma_3^+$ $(a, 0)$) of the two $I4/mcm$ phases were constrained to be the same except the symmetric strain $\Gamma_1^+$ which describes thermal expansion. Between 330 K and 380 K a two-model refinement was required using $Pbcm$ and $Cmcm$. All peaks in the XRD data were indexed based on the pseudocubic $Pm\bar{3}m$ phase to allow easy comparison.

The as-synthesized ceramic powder was prepared for TEM investigations through grinding and mixing with aluminium powder in a ratio of approx. 1:4, followed by cold rolling in a wrap of aluminium foil. This created a thin sheet of metal containing isolated ceramic particles, suitable for further mechanical thinning. A piece of this material was mounted on a platen with thermoplastic wax and ground with SiC paper (#1200 and #4000 grit) to a thickness of less than 30 µm, before a 2mm hole Cu support was attached with epoxy resin, and the sample was released from the platen. Electron transparency was obtained by $Ar^+$ ion milling (Gatan PIPS) at 6 keV, finishing at 0.5 keV. TEM investigations (bright field BF, dark field DF, and selected area electron diffraction SAED) were carried out in JEOL 2100 and JEOL 2100plus $LaB_6$ transmission electron microscopes at 200 kV. *In-situ* heating and cooling was carried out using a Gatan 652 furnace holder and a Gatan 636 liquid nitrogen cooling holder, respectively. In TEM/ED crystallographic indices corresponding to the aristotype structure are used, where the subscript *pc* denotes a pseudocubic unit cell.

*R.S. and C.A.C. contributed equally to this work
†Contact author: R.Beanland@warwick.ac.uk
‡Contact author: M.Senn@warwick.ac.uk

Table S1 provides details on all relevant $TiO_6$ distortions (tilting and displacive), including irrep label and OPD, relevant space group, **k** vector, ED reflection condition, and zone axis. [37]

## III. RESULTS

At low temperatures, our structural investigation by PXRD agrees with previous work, i.e. the material exhibits the $a^-a^-c^{++--}$ $Pbcm$ crystal structure, which is related to the $Pm\bar{3}m$ aristotype by a $\{(1,1,0),(\bar{1},1,0),(0,0,4)\}$ transformation. The Rietveld fit at 100 K is presented in Supplementary Information Figure S1,[37] while a magnified version highlighting specific reflections relating to the four different OPs, $T_2$ $\Delta_5$, $R_5^-$, and $M_5^-$ is shown in Figure 2(a). Peak splitting indicating the presence of a small tetragonal distortion (0.05 %), given by the $\Gamma_3^+(a,0)$ mode, and a lack of shear strain, (the $\Gamma_5^+(0,0,a)$ mode), are also shown in Figure 2(b) and (c), respectively. Given the lack of anticipated $\Gamma_5^+$ shear strain it is tempting to suggest a tetragonal $P4/nbm$ symmetry (which supports $T_2(0,0;a,0;0,0)$ and $R_5^-(0,0,b)$) rather than $Pbcm$. However, as shown in Supplementary Information Figure S2,[37] this would infer a slight shift to the R-point reflection (due to the different associated OPD) confirming that $Pbcm$ is the correct low temperature phase.

Figure 3 shows the RT $Pbcm$ microstructure as seen in a crystal in $[100]_{PC}$ orientation in the transmission electron microscope. The ¼ order Δ-point superstructure reflections along $001_{PC}$ are clearly visible in the $[100]_{PC}$ SAED pattern, Figure 3(b). By selecting one of these reflections a dark field (DF) Δ-point image can be produced as shown in Figure 3(a). Two types of domains become visible: bright regions show material that contains the $\Delta_5$ mode, patterned with fine black zig-zag lines with directions close to $[010]_{PC}$; and darker regions, where Δ-point reflections are absent. While this microstructure has previously been interpreted as a phase mixture,[30–33] a more straightforward interpretation is that this simply shows different orientational (pseudo-merohedral twin) $Pbcm$ domains, with the dark domains in $[001]_{PC}$ orientation. Figure 3(c) shows the diffraction pattern from these domains have no Δ-point superstructure reflections but do have *M*-point superstructure reflections, as expected for the $Pbcm$ structure with $M_5^-$ SOJT displacements. The domain walls separating these orientational variants have no preferred crystallographic direction, indicating a very low ferroelastic strain, in agreement with the small $\Gamma_3^+$ strain determined by PXRD.

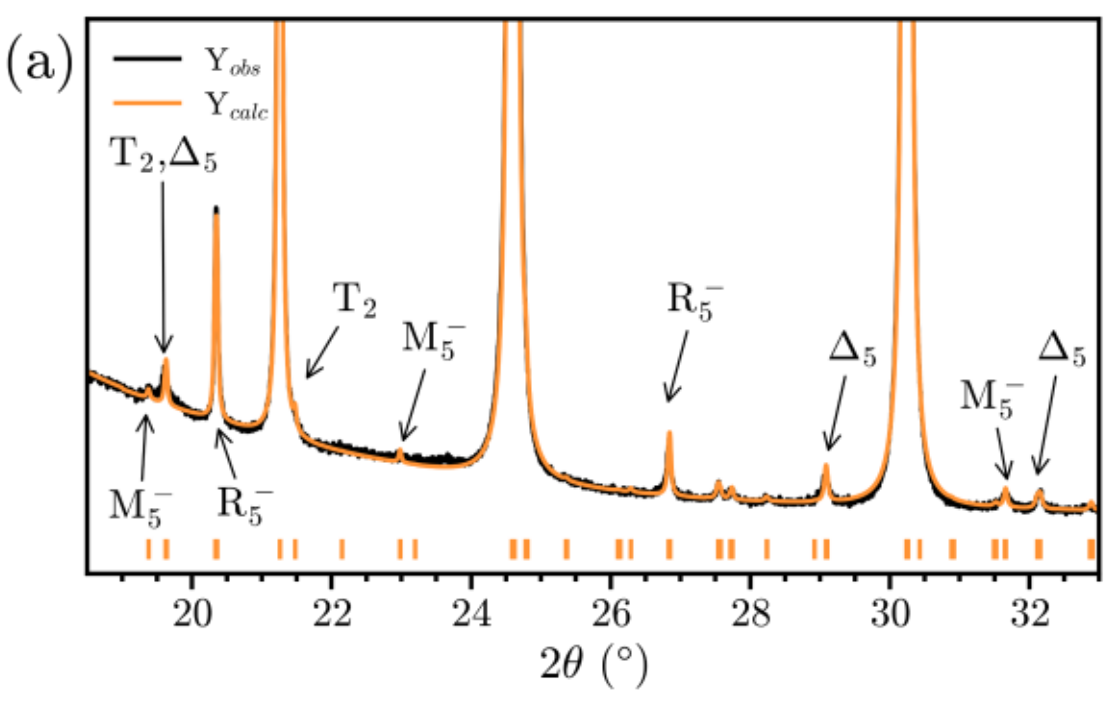

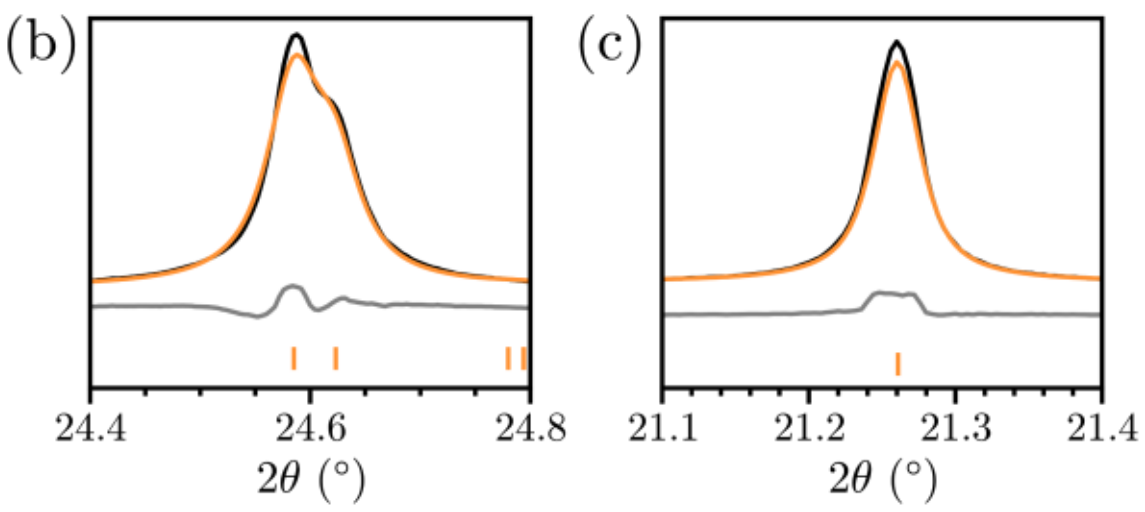

*FIG 2. PXRD data from $Ca_{0.4}Sr_{0.6}TiO_3$ at 100 K, with associated fit (see also Supplementary Information Figure S1. (a) Selected superstructure peaks identified according to their OPs. (b) Peak splitting in the ($200_{PC}$/$002_{PC}$) peak, evidencing small amount of $\Gamma_3^+(a,0)$ distortion. (c) Lack of splitting of ($111_{PC}$) /(1-$11_{PC}$), indicating negligible shear distortion, $\Gamma_5^+$*

The fine black lines in the bright $[100]_{PC}$ domains are planar defects with a preferred $(001)_{PC}$ habit plane. They indicate interruptions in the AFD $\Delta_5$ mode, which consists of displacements of Ti towards the edge of $TiO_6$ octahedra. The contrast is consistent with stacking faults characterised by a $\mathbf{R}=\frac{1}{2}\langle 100\rangle$ fault vector, i.e. anti-phase boundaries equivalent to an additional in-phase tilted layer of octahedra. Given that by symmetry analysis we know the $T_2$ and $\Delta_5$ order parameters are coupled, the observation of stacking faults in $\Delta_5$ simply imply corresponding anti-phase boundaries that exist in the tilting. The image therefore shows that the coherence of the $T_2$ mode is quite limited along the $[001]_{PC}$ direction, tens of nm at most and often much less, while it can extend for hundreds of nm perpendicular to the *c*-axis in the $(001)_{PC}$ plane. This reflects the relative ease of reversing the sequence of octahedral tilting about $[001]_{PC}$ with an antiphase boundary in the $(001)_{PC}$ plane, while the same boundary in a $(100)_{PC}$ orientation must have a wider transition and correspondingly higher energy.[34] The $R_5^-(a, a, 0)$ order parameter, producing corrugated anti-phase octahedral tilts is relatively robust in comparison. The resulting orientational anisotropy in energy thus drives the microstructure towards $(001)_{PC}$ sheets of coherent structure separated by planar defects. As pointed out by Howard[14] the near-cubic lattice parameter means these defects have relatively low energies.

*R.S. and C.A.C. contributed equally to this work
†Contact author: R.Beanland@warwick.ac.uk
‡Contact author: M.Senn@warwick.ac.uk

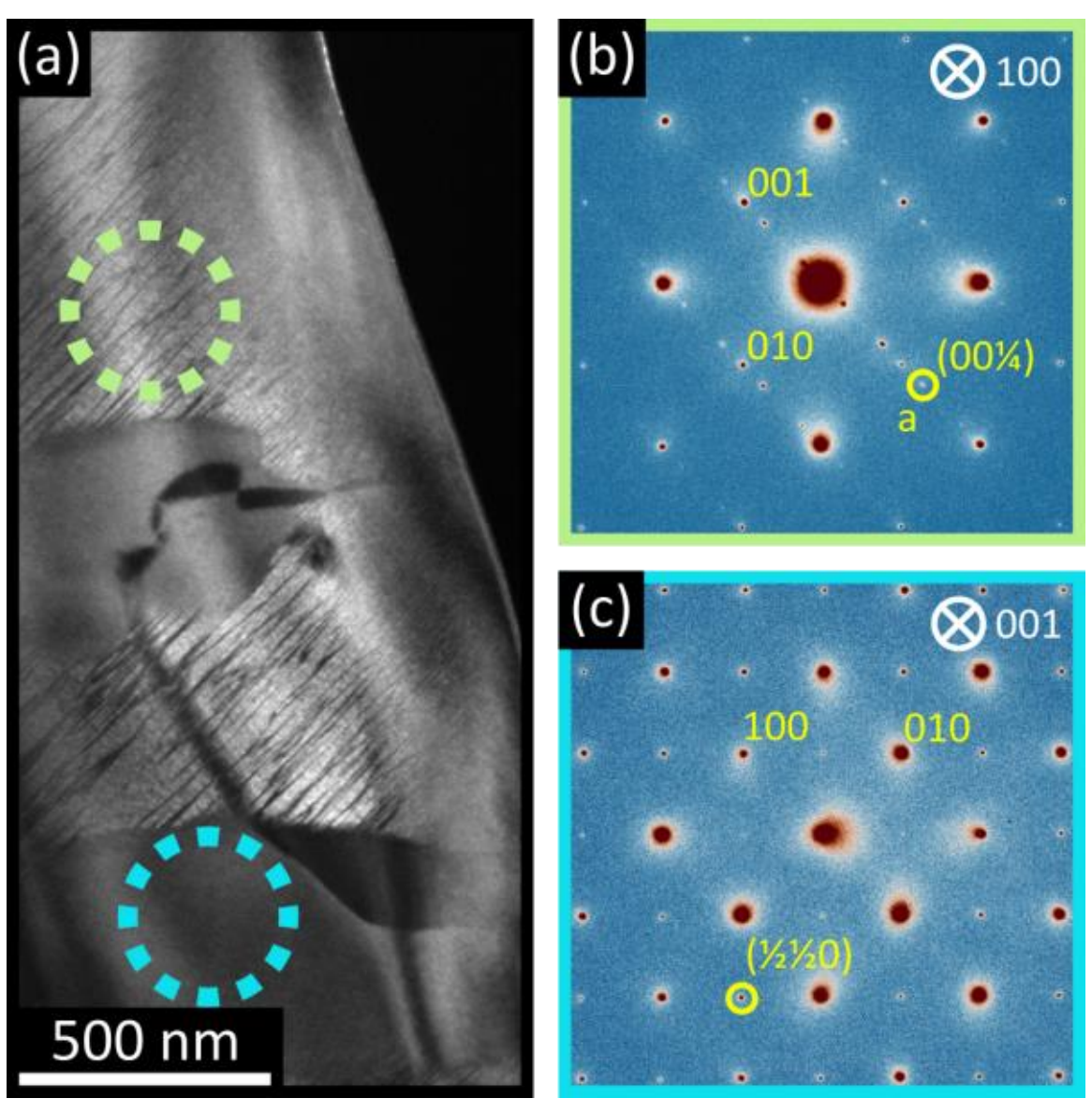

*FIG 3. TEM of room temperature $Pbcm$. (a) The DF TEM micrograph generated from a Δ-point reflection illustrates the presence of pseudo-merohedral twin domains, and antiphase boundaries perpendicular to the c-axis in $[100]_{PC}$ domains. The meandering black band is a planar defect that exists in both domains, presumably an anti-phase boundary affecting the $R_5^-$ antiphase octahedral tilt common to both. The <100> type electron diffraction patterns in (b) and (c) are recorded from the two regions indicated by the dashed circles. Indexing corresponds to the aristotype cubic $Pm\bar{3}m$ structure. False colour is used to emphasise the presence of weak superstructure reflections.*

Notably, since $M_2^+$ is absent in the $Pbcm$ phase, the $M$-point superstructure reflections observed at RT are not associated with in-phase octahedral rotations. Thus, $M$-point ½*ooe* superstructure spots clearly visible in the $[001]_{PC}$ SAED pattern of Figure 3(c) are associated with the $M_5^-$ SOJT distortion of $TiO_6$ octahedra, also observed in the PXRD data in Figure 2(a). Together, the X-ray and TEM data suggest that the AFD $M_5^-$ and $\Delta_5$ modes are coupled to, and perhaps driven by, the interplay between $T_2$ ↓↓↑↑ octahedral and $R_5^-$ octahedral tilts. The relatively small AFD distortion is consistent with the large degree of octahedral tilting, which reduces the Ti $d$ and O $p$ orbital overlap thereby destabilising a SOJT distortion.

The $Pbcm$ model accurately fits the variable temperature PXRD data up to about 330 K (see Supplementary Information Figure S3[37] for detailed refinement outputs). Magnitudes of octahedral tilting $T_2$, and AFD cation displacements $\Delta_5$, observed with PXRD and ED are shown in Figure 4. Δ-point and $T$-point superstructure reflections are visible in $[100]_{PC}$ and $[310]_{PC}$ SAED patterns respectively (Figures 4(a) and (b)) and both vanish, as expected at the phase transition (Figure 4(c)). The ED intensities are rendered rather unreliable by bending of the TEM specimen during heating but are generally consistent with the mode amplitudes extracted from Rietveld refinement of PXRD data (Figure 4(d)).

We note there is a discrepancy between what is measured by TEM and PXRD data (Figures 4(c) and 4(d) respectively) due to the varying precision of the temperature control, the sensitivities to order parameter coherence lengths, and also the fact that for TEM we are plotting the intensity of single reflection, subject strongly to dynamical effects, rather than the refined amplitudes of the order parameters. Measurement of Δ-point superstructure reflections in SAED patterns is less reliable due to dynamical coupling with adjacent strong reflections, and in this data, they have a maximum intensity at 200 K before diminishing and disappearing at the same temperature as $T$-point reflections.

*R.S. and C.A.C. contributed equally to this work
†Contact author: R.Beanland@warwick.ac.uk
‡Contact author: M.Senn@warwick.ac.uk

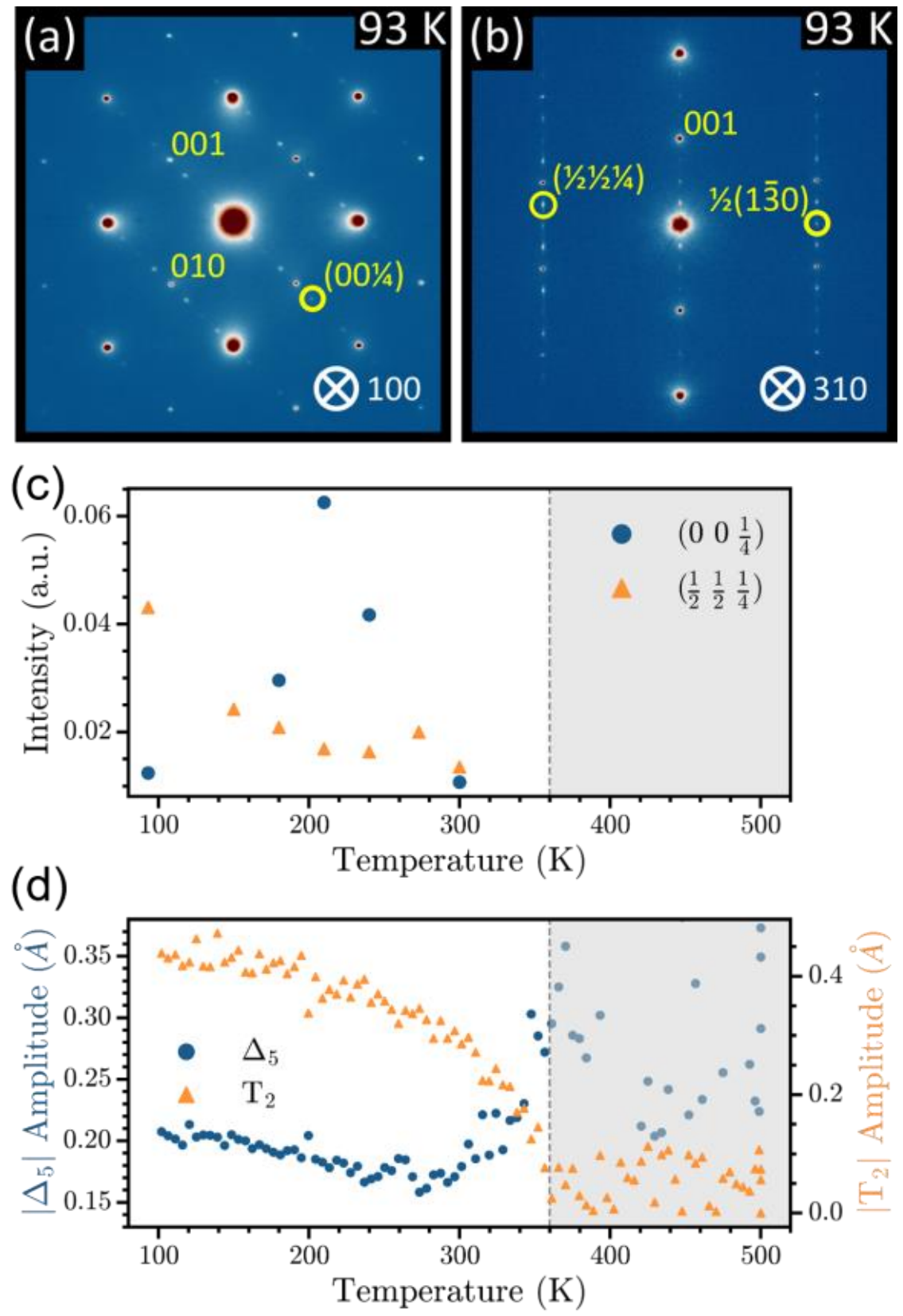

*FIG 4. The AFD Δ mode and c-axis T tilt mode in ED and PXRD as a function of temperature. In ED, $[100]_{PC}$ is used for Δ-point reflections (a), while $[310]_{PC}$ is used for T-point reflections (b). The summed intensity of all superstructure reflections are shown in (c), normalized to the summed intensity of all matrix reflections in the pattern. Mode amplitudes from PXRD data are given in (d). Shaded region above 360 K marks the transition from Pbcm to $I4/mcm$.*

The mode amplitudes plotted in Figure 4(d), show a gradual decrease in $T_2$, and corresponding decrease in $\Delta_5$, as temperature increases from 100 to 300 K. Above this point $T_2$ decreases rapidly and modelling of the AFD modes $\Delta_5$ and $M_5^-$ loses validity above 370 K, producing erratic amplitudes that are untethered by the PXRD data, (see also Figure S3[37]). However, before this point, between 290 and 360 K there is evidence that the $\Delta_5$ and $T_2$ modes compete just before reaching the structural phase transitions. In previous investigations it has been proposed that the emergence of the $\Delta_5$ OP in $Pbcm$ is due to the interaction between $R_5^-$ and $T_2$.[12] While there is some evidence in Figure S3(e) and (f) that the $\Delta_5$ Ti displacements evolve mirroring $T_2$, the overall trend of the magnitude of the two order parameters shown in Figure 4 (and Figure S3(e) and (g)), suggests some degree of competition between the octahedral tilting ($T_2$) and AFD SOJT ($\Delta_5$) modes indicating that they arise from distinct origins. Indeed, an analysis of the free energy invariant terms in the Landau-style expansion of these order parameters about parent $Pm\bar{3}m$ symmetry reveals two competing third order terms in the free energy: $\alpha(\Delta_5)^2(\Gamma_3^+) + \beta(T_2)^2(\Gamma_3^+)$. These both feature the same secondary order parameter, transforming as the symmetry-adapted tetragonal strain, $\Gamma_3^+$, whose derivative with temperature changes sign around 310 K (see Figure S3(c)[37]). We note a similar competitive coupling also exists with $X_1^+$, that, similarly to $\Delta_5$ modes, encode off-centre displacements of Ti cations, but with a competing periodicity. It thus seems there are two temperature regimes, one in which the third order coupling between $\Delta_5$, $R_5^-$, and $T_2$ dominate the free energy, and another characterised by the competition between third order terms such as $\alpha(\Delta_5)^2(\Gamma_3^+) + \beta(T_2)^2(\Gamma_3^+)$.

This competition resolves itself into a clear phase separation above 330 K, when heating from 100 K to 500 K as shown in Figure S4.[37] Figure S5 details the quality of fits to various combinations of different phases at 370 K. This 330 – 380 K phase transition is best fit by the following sequence: $Pbcm \rightarrow (Pbcm + Cmcm) \rightarrow (I4/mcm + I4/mcm)$, where an additional $Cmcm$ phase has a far better fit compared to an additional $Pnma$, $I4/mcm$, or a second $Pbcm$ phase (see Figure S5). The intermediate $Cmcm$ phase exhibits a different distortion of lattice parameters compared to the $Pbcm$ phase. This is due to the similarities in the $Cmcm$ $\Gamma_3^+$ strain OPD, $(a, b)$, which closer to that present in $I4/mcm$, $(a, 0)$.

This phase separation persists above the tetragonal phase transition at 380 K, resulting in two distinct $I4/mcm$ phases shown in Figures S6 and S7, that are identical in all aspects, excepting their symmetric strains, $\Gamma_1^+$, i.e. unit cell volumes, which differ by 0.2 % (Figure S8[37]). The merging of Ti-O bond angles above 300 K (Figure S9[37]) due to the competing coupling terms in turn allows for a greater variety of displacive directions to emerge, which we speculate may be one possibility driving the phase separation into $Pbcm$ and $Cmcm$, however this effect is still not fully understood at present. This degree of strain between the two $I4/mcm$ phases is well below the detection limit of TEM imaging or electron diffraction, and in addition to the competing distortions, changes to the domain structure may also contribute to this phase separation as discussed below.

*R.S. and C.A.C. contributed equally to this work
†Contact author: R.Beanland@warwick.ac.uk
‡Contact author: M.Senn@warwick.ac.uk

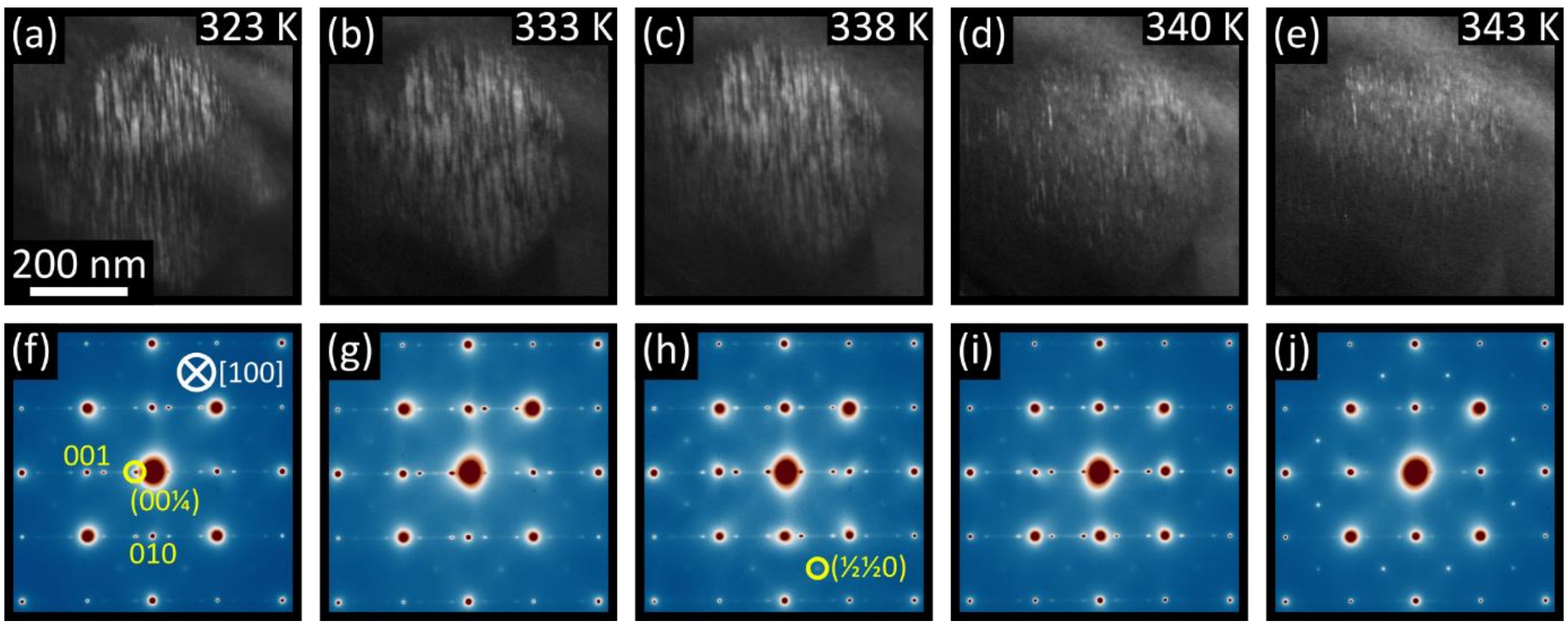

*FIG 5. In-situ heating of $[100]_{PC}$ $Ca_{0.4}Sr_{0.6}TiO_3$ in TEM. Top row (a)-(e): DF images taken using a* $\mathbf{k}$ *= (0 0 ¼) Δ-point superstructure reflection. Bottom row (f)-(j): the corresponding $[100]_{PC}$ diffraction patterns. The vertical black stripes show faults in the stacking of the $Pbcm$ phase and show subtle changes up to 338 K. At 340 K they rapidly expand and merge, leaving only small planar regions of the $Pbcm$ phase. As the Δ-point reflections show a corresponding diminishing intensity,* $\mathbf{k}$ *= (½ ½ 0) M-point reflections appear and gain intensity (see circle in (h)).*

It is worth noting that when this material is heated from 300-500 K without prior cooling (to 100 K), the peak splitting evidencing the $Pbcm$ and $Cmcm$ phase separation or two $I4/mcm$ phases was not present using PXRD (see Figure S10[37]). This result hence highlights just how sensitive these samples will be to their thermal history, and great care should be taken when drawing comparisons across the literature.

To further investigate this change in behaviour with thermal history, the behaviour of Δ-point superstructure reflections was examined on heating from RT. The corresponding microstructural changes were also investigated using VT dark field imaging, as this AFD mode is lost during the transition from $Pbcm$ to $I4/mcm$. Figure 5(a) shows a Δ-point DF image of a small domain with orientation $[100]_{PC}$ and its corresponding SAED pattern in Figure 5(f) at 323 K. The Δ-point superstructure reflections arise from bright bands a few nm wide lying perpendicular to $(001)_{PC}$, with a high density of anti-phase boundaries, similar to those in Figure 3(a), seen as dark vertical lines. Only subtle changes in structure are observed until 338 K (Figures 5(b), (c), (g) and (h)), above which the bright regions rapidly begin to become thinner while the dark bands become wider, indicating increasing loss of coherence along $[001]_{PC}$. Above 340 K the microstructure becomes finer, with many isolated, extremely narrow $(001)_{PC}$ sheets, with thickness of only a few nm, which contract laterally with increasing temperature (Figures 5(d), (e)). Simultaneously, the Δ-point superstructure reflections decrease in intensity, (Figures 5(i), (j)) fading into continuous streaks along $001_{PC}$ as coherence is lost. These images show that the behaviour of the $\Delta_5$ OP in Figure 4(d) above 300 K is due to regions that start to fall below the detection limit of PXRD as the phase transition is approached and their dimensions approach the nanoscale.

Interestingly as superstructure reflections due to $T_2$ and $\Delta_5$ fade away, $M$-point superstructure reflections appear in the $[100]_{PC}$ SAED pattern, as highlighted in Figure 5(h). These reflections should not exist in a material with the $a^0a^0c^-$ $I4/mcm$ structure (and indeed are not detected in PXRD regardless of thermal history, indicating that they have very short coherence lengths). Electron microscopy conducted in the $[112]_{PC}$ direction (of a sample which had not been prior cooled), where both in-phase tilting and out-of-phase tilting reflections can be observed simultaneously, was used to confirm the retention of the out-of-phase $R_5^-$ (**k** = (½ ½ ½)) reflections and the diminishing intensity of the *M*-point (**k** = (½ ½ 0)) reflections after the phase transition above 400 K, as shown in Figure S11.[37] These observations confirm that even without additional cooling, that the phase transition from $Pbcm$ to $I4/mcm$ is rather more complex, and suggests that the higher temperature phase contains nanoscale regions that are quite different to the average structure, at least in the vicinity of the phase boundary.

*R.S. and C.A.C. contributed equally to this work
†Contact author: R.Beanland@warwick.ac.uk
‡Contact author: M.Senn@warwick.ac.uk

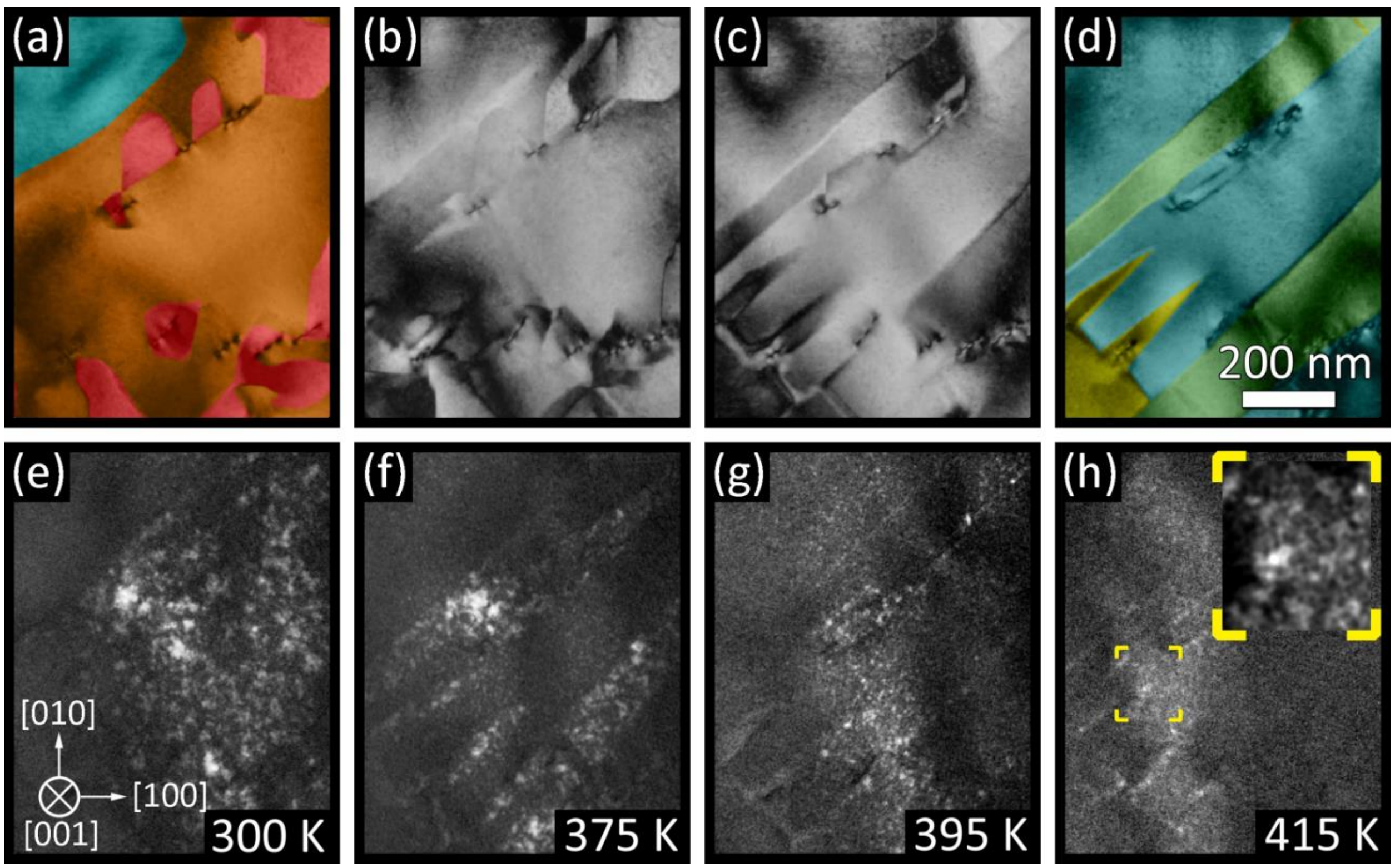

*FIG 6. In-situ heating of $[001]_{PC}$ $Ca_{0.4}Sr_{0.6}TiO_3$ in TEM (see Figure S12 for corresponding diffraction patterns [37]). Bright field images (top) show the change in domain structure through the phase transition. False colour is used in the initial and final images to highlight domain structure; the top left (teal) region in (a) is in $[100]_{PC}$ orientation. Dark field, M-point images (bottom) show the origin of intensity in **k** = (½½0) type reflections.*

To investigate this further, $M$-point superstructure reflections were examined across the phase transition in a region containing $[001]_{PC}$ domains (Figure 6). (Part of the crystal in Figure 6 is a $[100]_{PC}$ domain giving similar information to that shown in Figure 5, see Figure S11 for additional information) Bright field images, Figures 6(a)-(d) show the evolution of domain structure from 300 K – 415 K. Figure 6(d) shows that the high temperature phase has straight domain walls on $\{110\}_{PC}$ planes, whereas domain walls in the low temperature $Pbcm$ structure have no strong orientational preference in Figure 6(a). The complete lack of correlation in the positions of domain walls indicates a phase transition with no group-subgroup relationship, as expected. Diffraction patterns contain **k** = (½ ½ 0) $M$-point superstructure spots throughout the temperature range (Figure S12[37]) and the DF $M$-point images Figures 6(e)-(h) show the regions that contribute to them. At RT, Figure 6(e), these images show the AFD $M_5^-$ OP present in $Pbcm$, which appears as mottled contrast. In this $[001]_{PC}$ view the stacking faults in the $T_2$ ↓↓↑↑ OP, seen edge-on in Figures 3 and 5, lie in the plane of the specimen. Because they are perpendicular to the direct beam, they are invisible in this orientation but introduce phase shifts in the coupled $M_5^-$ OP. As a result, the electron beam also experiences several phase shifts in its passage through the specimen, resulting in constructive or destructive interference that depends upon the number of faults and their location. This is the origin of the mottled contrast in the $M$-point images. Importantly for CBED measurements of local symmetry, the phase shifts produced by these faults break the symmetry of the specimen with respect to the electron beam, change extinction conditions and are almost certainly responsible for the incorrect $P2_12_12$ space group determined previously.

As the temperature increases, a phase coexistence is observed, for example in Figure 6(f) at 375 K, where much of the mottled area has been replaced by darker bands that have many small nanoscale regions of bright contrast. These regions decrease in size as temperature is increased (Figures 6(g) and (h). This behaviour is the same as that shown in Figure 5, but here seen from a different direction. As shown in the inset to Figure 6(h), well above the phase transition at 415 K, these bright regions are very small, only a few nm across, but still give easily detectable intensity in $M$-point superstructure spots (Figure S12[37]). Three additional features observable in these data are worth noting. First, weak streaks along $001_{PC}$-type directions are observed in the diffraction patterns at 395 K and 415 K (Figure S12). These indicate the presence of very thin platelets of material on $(001)_{PC}$ planes. The appearance of these streaks in all patterns at 415 K indicates that thin platelets of different structure form with all $\{001\}_{PC}$ orientations. Second, the $M$-point superstructure spots show a pattern of weak intensities along $110_{PC}$ consistent with systematic absences (Figures 5(j) and S12). Together, these observations suggest that these platelets have in-phase $M_2^+$ tilting. Third, while the origin of these superstructure spots spreads out to cover all domains in Figures 6(g) and (h), there is a significantly higher density at domain walls, suggesting that the lower symmetry that must locally exist at the wall[35] lowers the energy of formation for these platelets.

*R.S. and C.A.C. contributed equally to this work
†Contact author: R.Beanland@warwick.ac.uk
‡Contact author: M.Senn@warwick.ac.uk

In summary, while the low temperature $Pbcm$ phase is characterised by slabs of material that extend for hundreds of nm perpendicular to the $c$-axis, the $a^0a^0c^-$ $I4/mcm$ phase found above 380 K contains a high density of nanoscale platelets on $\{100\}_{PC}$ planes, only a few unit cells in thickness and a few nm wide. The $M$-point superstructure reflections associated with them suggest they are small sheets of material with some in-phase tilting ($M_2^+$). Such nanoscale sheets, with octahedral tilting that differs from the macroscopic structure, have been observed before in perovskite oxides, notably in $Na_{0.5}Bi_{0.5}TiO_3$.[34,36] Here, their presence can be understood as a response to the statistical fluctuations in composition – that must occur in a compound material at an atomic scale – in combination with the very steep change in phase transition temperature with composition (roughly 14 K for d$x$ = 0.01, i.e. $dT_{C2}/dx \sim 1400\ K$). We may thus understand Figure 6 to show the presence of nanoscale regions where a second OP induces octahedral tilts in addition to the dominant OP $a^0a^0c^-$ $R_5^-(0,0,a)$ close to the phase transition temperature. The gradual transition from slabs of coherent structure into nanoscale platelets seen in Figure 5 can also be understood as a consequence of the large $dT_{C2}/dx$, with slightly Ca-rich regions losing coherence first, while others maintain remnants of the $T_2$ ↓↓↑↑ octahedral tilts with progressively finer length scales as the temperature increases. As noted by Howard et al.[14] a mix of in-phase and antiphase tilts corresponds to modes along the T line of symmetry, **k** = (½ ½ ξ). They also noted that such a structure should give rods of diffuse intensity along $001_{PC}$-type directions in reciprocal space at positions **g** + ½ ½ ξ, where **g** is an aristotype reciprocal lattice vector, precisely as observed in Figure S12.

This nanoscale microstructure, invisible to PXRD, explains the Raman measurements that indicate oxygen octahedral tilting incompatible with the nominal (macroscopic) $I4/mcm$ structure.[18] Indirect evidence, in the form of softening of elastic constants extending to at least 50 K above the phase transition, has also been observed by Manchado et al.[20] studied the first order $I4/mcm$ to $Pnma$ phase transition in $Ca_xSr_{1-x}TiO_3$ using calorimetry, dielectric measurements and resonant ultrasound measurements. They showed that hysteresis in the phase transition temperature $T_{C2}$ of 20 K or more is typical, which agrees with our observations (Figures 5 and 6). We find the nature of this phase transition also has an impact on measurements made in sequence. When cooling from $I4/mcm$ to $Pbcm$, it is found that the slabs of coherent structure are much thinner than in that found in material that has not been heated or cooled beforehand. This can be seen by comparing Figure 5, in which the material has already been cycled through the phase transition, and Figure 3, which has not received any thermal treatment. Indeed, the presence of anti-phase boundaries can be seen as a result of cooling through $T_{C2}$ with many independent nucleation centres for the $Pbcm$ phase that are provided by the nanoscale sheets.

$Ca_{0.4}Sr_{0.6}TiO_3$ clearly shows microstructural changes on heating from RT to above the $I4/mcm$ phase transitions, where short range in-phase tilting occurs that is not evident in the PXRD data on heating with the same thermal history. We further investigated this by precooling a TEM specimen to 80 K (in liquid $N_2$), prior to a RT TEM experiment. Within the specimen at RT, we observe grains with the ¼ order spots characteristic of $Pbcm$, and we additionally observe grains without these reflections, and instead show ½ *eeo* spots (Figure S13[37]) which are often used to identify systems with tilting about multiple axes. The domains imaged with these spots are still small, on the sub-50 nm scale, and at RT are unlikely to be observed by PXRD. However, these reflections would be consistent with the $Cmcm$ phase observed above 330 K in the PXRD data. This reflects the (micro)structural

*R.S. and C.A.C. contributed equally to this work
†Contact author: R.Beanland@warwick.ac.uk
‡Contact author: M.Senn@warwick.ac.uk

complexities that this material exhibits with varying temperature.

## IV. CONCLUSIONS

We have studied the phase transition behaviour of $Ca_{0.4}Sr_{0.6}TiO_3$ across a wide temperature range, using a complementary combination of PXRD, TEM and electron diffraction (ED) techniques. As expected, the structure of the low temperature $Pbcm$ phase can be described using two primary order parameters, $R_5^-$ and $T_2$, describing antiphase $TiO_6$ octahedral tilts about $[110]_{PC}$ and a ↓↓↑↑ tilt pattern along $[001]_{PC}$ respectively, $a^-a^-c^{++--}$. Two further OPs describing antiferrodistortive displacements, $M_5^-$ and $\Delta_5$, are also present. All of these modes produce superstructure reflections that are observed in both PXRD and ED patterns, and they can be used in TEM to form dark field images that show the microstructure of each. There is some evidence that the $\Delta_5$ antiferrodistortive and $T_2$ tilting modes compete with each other as the phase transition and separation around 320 K is reached on warming from 100 K, leading to a rich microstructure that we have probed by TEM and model as a phase coexistence in our PXRD data. The higher temperature $I4/mcm$ phase only exhibits $R$-point reflections in PXRD patterns, consistent with an $R_5^-(0,0,a)$ mode describing antiphase $TiO_6$ octahedral tilts about $[001]_{PC}$ alone, $a^0a^0c^-$. However, in ED additional $M$-point reflections are readily observed, indicating the presence of in-phase tilts. The difference between the two measurements is a result of the smaller wavelength and stronger interaction of electrons, providing a sensitivity to nanoscale structures that evades X-ray diffraction, primarily due to Scherrer broadening of weak superstructure peaks. Dark field $\Delta$-point and $M$-point TEM images show that the phase transition proceeds by a variable loss of coherence in tilts along $[001]_{PC}$, leading to a microstructure of thin platelets with nanoscale dimensions along the $c$-axis. We propose that this behaviour is a result of the rapid change in phase transition with temperature ($dT_{C2}/dx \sim 1400\ K$) in combination with the statistical fluctuations of composition that are inevitable at the atomic scale in a compound material. Some of these platelets persist well above the phase transition temperature. As predicted by Carpenter[12] they produce rods of diffuse intensity along $001_{PC}$-type directions in reciprocal space.

The microstructure associated with the $Pbcm$ – $I4/mcm$ phase transition also manifests in several indirect ways, being responsible for hysteresis in the phase transition temperature and quite probably softening of elastic constants in the $I4/mcm$ phase. We find a significant distortion of lattice parameters when passing through the phase transition for a second time, and that within the transition region, the system is best fit by a coexistence of $Pbcm$ with a $a^0b^+c^-$ $Cmcm$ phase, again suggesting that a mix of tilt system exists even though no superstructure peaks are observed.

The discrepancy between Raman and diffraction measurements in this materials system has been a long-standing puzzle. X-ray diffraction unambiguously shows the $I4/mcm$ phase to be tetragonal,[13] while Raman spectra show the same phase to have orthorhombic symmetry or lower, with equal certainty.[16] While the idea that the local symmetry sampled by Raman spectroscopy is lower than the average macroscopic symmetry has been proposed for some time,[14] the nanoscale structures that might be responsible have remained hidden until now. More broadly, we expect some of the mechanisms elucidated here to be found in other compound perovskite oxide systems, particularly those in which there are large differences in phase transition temperatures of the end point materials.

## ACKNOWLEDGMENTS

We would first like to thank Z. Yang Lee for sample preparation including synthesis and initial characterisation, performed using equipment provided by the Warwick X-ray Research Technology Platforms. We thank the Warwick Electron Microscopy Research Technology Platform for use of equipment for TEM measurements. We additionally thank Diamond Light Source under the Oxford-Warwick BAG (CY39378-1, CY25166-13). We thank E. Bindoff and K. Gadher for assistance with TEM and ED. M.S. S acknowledges funding from the Royal Society (UF160265 and URF231012). R.B, R.S. and C.A.C acknowledge funding from the EPSRC grants "New directions in high temperature dielectrics: unlocking performance of doped tungsten bronze oxides through mechanistic understanding," and "High permittivity, quasi linear dielectrics for high field/temperature capacitors in power electronics".

*R.S. and C.A.C. contributed equally to this work
†Contact author: R.Beanland@warwick.ac.uk
‡Contact author: M.Senn@warwick.ac.uk

[1] Gränicher, H. & Jakits, O. Über die dielektrischen Eigenschaften und Phasenumwandlungen bei Mischkristallsystemen vom Perowskittyp. *Nuovo Cim* **11**, 480–520 (1954).
[2] McQuarrie, M. Structural Behavior in the System (Ba, Ca, Sr)TiO $_3$ and Its Relation to Certain Dielectric Characteristics. *J. Am. Ceram. Soc.* **38**, 444–449 (1955).
[3] Mitsui, T. & Westphal, W. B. Dielectric and X-Ray Studies of $Ca_xBa_{1-x}TiO_3$ and $Ca_xSr_{1-x}TiO_3$. *Phys. Rev.* **124**, 1354–1359 (1961).
[4] Redfern, S. A. T. High-temperature structural phase transitions in perovskite ($CaTiO_3$). *J. Phys.: Condens. Matter* **8**, 8267–8275 (1996).
[5] Lee, W. E., Ojovan, M. I. & Jantzen, C. M. *Radioactive Waste Management and Contaminated Site Clean-up: Processes, Technologies and International Experience*. (Woodhead Publishing, 2013).
[6] Reaney, I. M. *et al.* The role of chemical, polar and octahedral tilt disorder in high voltage/energy density ceramics. *Int. Mater. Rev.* **71**, 178–196 (2026).
[7] Müller, K. A. & Burkard, H. $SrTiO_3$: An intrinsic quantum paraelectric below 4 K. *Phys. Rev. B* **19**, 3593–3602 (1979).
[8] Shin, D. *et al.* Quantum paraelectric phase of $SrTiO_3$ from first principles. *Phys. Rev. B* **104**, L060103 (2021).
[9] Palani, P., Fasquelle, D. & Tachafine, A. A review on (Sr,Ca)$TiO_3$-based dielectric materials: crystallography, recent progress and outlook in energy-storage aspects. *Mater Sci* **57**, 12279–12317 (2022).
[10] Ranjan, R. & Pandey, D. Antiferroelectric phase transition in $(Sr_{1-x}Ca_x)TiO_3$: II. X-ray diffraction studies. *J. Phys. Condens. Matter* **13**, 4251–4266 (2001).
[11] Ranjan, R., Pandey, D. & Lalla, N. P. Novel Features of $Sr_{1-x}Ca_xTiO_3$ Phase Diagram: Evidence for Competing Antiferroelectric and Ferroelectric Interactions. *Phys. Rev. Lett.* **84**, 3726–3729 (2000).
[12] Carpenter, M. A., Howard, C. J., Knight, K. S. & Zhang, Z. Structural relationships and a phase diagram for (Ca,Sr)$TiO_3$ perovskites. *J. Phys.: Condens. Matter* **18**, 10725–10749 (2006).
[13] Howard, C. J. *et al.* Space-group symmetry for the perovskite $Ca_{0.3}Sr_{0.7}TiO_3$. *J. Phys.: Condens. Matter* **17**, L459–L465 (2005).
[14] Howard, C. J., Withers, R. L., Knight, K. S. & Zhang, Z. $(Ca_{0.37}Sr_{0.63})TiO_3$ perovskite—an example of an unusual class of tilted perovskites. *J. Phys.: Condens. Matter* **20**, 135202 (2008).
[15] Mishra, S. K., Ranjan, R., Pandey, D. & Kennedy, B. J. Powder neutron diffraction study of the antiferroelectric phase transition in $Sr_{0.75}Ca_{0.25}TiO_3$. *J. Appl. Phys.* **91**, 4447–4452 (2002).
[16] Mishra, S. K., Ranjan, R., Pandey, D. & Stokes, H. T. Resolving the controversies about the 'nearly cubic' and other phases of $Sr_{1-x}Ca_xTiO_3$ ($0 \leq x \leq 1$): I. Room temperature structures. *J. Phys.: Condens. Matter* **18**, 1885–1898 (2006).
[17] Ranjan, R., Pandey, D., Siruguri, V., Krishna, P. S. R. & Paranjpe, S. K. Novel structural features and phase transition behaviour of $(Sr_{1-x}Ca_x)TiO_3$ : I. Neutron diffraction study. *J. Phys.: Condens. Matter* **11**, 2233–2246 (1999).
[18] Mishra, S. K. *et al.* A combined X-ray diffraction and Raman scattering study of the phase transitions in $Sr_{1-x}Ca_xTiO_3$ (x = 0.04, 0.06, and 0.12). *J. Solid State Chem.* **178**, 2846–2857 (2005).
[19] Ranson, P. *et al.* The various phases of the system $Sr_{1-x}Ca_xTiO_3$ - A Raman scattering study. *J. Raman Spectrosc.* **36**, 898–911 (2005).
[20] Manchado, J. *et al.* Dielectric, calorimetric and elastic anomalies associated with the first order I4/mcm → Pbcm phase transition in (Ca, Sr)$TiO_3$ perovskites. *J. Phys.: Condens. Matter* **21**, 295903 (2009).
[21] Glazer, A. M. The classification of tilted octahedra in perovskites. *Acta Cryst. B* **28**, 3384–3392 (1972).
[22] Howard, C. J. & Stokes, H. T. Group-Theoretical Analysis of Octahedral Tilting in Perovskites. *Acta Cryst. B* **54**, 782–789 (1998).
[23] Senn, M. S. & Bristowe, N. C. A group-theoretical approach to enumerating magnetoelectric and multiferroic couplings in perovskites. *Acta Cryst. A* **74**, 308–321 (2018).
[24] Megaw, H. D. The seven phases of sodium niobate. *Ferroelectrics* **7**, 87–89 (1974).
[25] Shuvaeva, V. A. *et al.* Crystal structure of the electric-field induced ferroelectric phase of $NaNbO_3$. *Ferroelectrics* **141**, 307–311 (1993).
[26] Campbell, B. J., Evans, J. S., Perselli, F. & Stokes, H. T. Rietveld refinement of structural distortion-mode amplitudes. *IUCr Comput. Comm. Newsl.* 81–95 (2007).

*R.S. and C.A.C. contributed equally to this work
†Contact author: R.Beanland@warwick.ac.uk
‡Contact author: M.Senn@warwick.ac.uk

[27] Coelho, A. A. *TOPAS* and *TOPAS-Academic*: an optimization program integrating computer algebra and crystallographic objects written in C++. *J. Appl. Crystallogr.* **51**, 210–218 (2018).
[28] Stokes, H. T., Hatch, D. M. & Campbell, B. J. ISODISTORT, ISOTROPY Software Suite, iso.byu.edu. iso.byu.edu.
[29] Campbell, B. J., Stokes, H. T., Tanner, D. E. & Hatch, D. M. *ISODISPLACE* : a web-based tool for exploring structural distortions. *J. Appl. Crystallogr.* **39**, 607–614 (2006).
[30] Anwar, S. & Lalla, N. P. Electron microscopic studies of the antiferroelectric phase in $Sr_{0.60}Ca_{0.40}TiO_3$ ceramic. *J. Solid State Chem.* **181**, 997–1004 (2008).
[31] Anwar, S. & Lalla, N. P. Occurrence of a new superlattice phase across the antiferroelectric phase transition in $Sr_{1-x}Ca_xTiO_3$ (x = 0.30 and 0.40). *J. Phys.: Condens. Matter.* **20**, 325231 (2008).
[32] Anwar, S. & Lalla, N. P. Phase coexistence in $Sr_{0.70}Ca_{0.30}TiO_3$ studied through electron diffraction. *Solid State Sci.* **10**, 307–315 (2008).
[33] Anwar, S. & Lalla, N. P. Space group analysis of $Sr_{1-x}Ca_xTiO_3$ ceramics with $x$ = 0.20, 0.27 and 0.30 through electron diffraction. *J. Phys.: Condens. Matter.* **19**, 436210 (2007).
[34] Beanland, R. Structure of planar defects in tilted perovskites. *Acta Cryst. A* **67**, 191–199 (2011).
[35] Salje, E. K. H. Multiferroic Domain Boundaries as Active Memory Devices: Trajectories Towards Domain Boundary Engineering. *ChemPhysChem* **11**, 940–950 (2010).
[36] Beanland, R. & Thomas, P. A. Imaging planar tetragonal sheets in rhombohedral $Na_{0.5}Bi_{0.5}TiO_3$ using transmission electron microscopy. *Scr. Mater.* **65**, 440–443 (2011).
[37] See Supplemental Material for descriptions of distortion modes, Rietveld refinements and extracted structural information, and additional electron diffraction and microscopy.

*R.S. and C.A.C. contributed equally to this work
†Contact author: R.Beanland@warwick.ac.uk
‡Contact author: M.Senn@warwick.ac.uk